\documentclass{fundam}

\usepackage[utf8]{inputenc}
\usepackage{graphicx}
\usepackage{amssymb}
\usepackage{amsmath}
\usepackage{url}

\newcommand{\rys}[1]{}
\newcommand{\citep}{\cite}
\newcommand{\wyrzucicv}[1]{\textbf{}}

\newcommand{\Bem}[1]{}
\newcommand{\HEM}{\mathcal{E}}
\newcommand{\hem}{\mathfrak{e}}
\newcommand{\HGM}{\mathcal{H}}
\newcommand{\N}{ V }
\newcommand{\n}{ v }

\newcommand{\HG}{H}
\newcommand{\HE}{E}
\newcommand{\GG}{\mathcal{G}}
\newcommand{\GE}{\mathbf{E}}
\newcommand{\Ge}{\mathbf{e}}
\newcommand{\GN}{\mathbf{V}}

\newcommand{\Zeta}{\boldsymbol\zeta}

\begin{document}

\setcounter{page}{49}
\publyear{22}
\papernumber{2151}
\volume{189}
\issue{1}

\finalVersionForARXIV

\title{Network Capacity Bound for Personalized  PageRank in Multimodal Networks}

\author{Mieczys{\l}aw A. K{\l}opotek\thanks{Address for correspondence: Institute of Computer Science,
                  Polish Academy of Sciences, Warsaw, Poland.  \newline \newline
          \vspace*{-6mm}{\scriptsize{Received October 2022; \ accepted December 2022.}}},  S{\l}awomir  T. Wierzcho{\'n},
            Robert  A. K{\l}opotek
     \\
Institute of Computer Science \\
Polish Academy of Sciences, Warsaw, Poland\\
klopotek@ipipan.waw.pl
}

\maketitle

\runninghead{M.A. K{\l}opotek et al.}{Multimodal Network Pagerank Bound}

\begin{abstract}
\noindent
In a former paper \cite{Bipartite:2016}  the concept of Bipartite PageRank was introduced and a theorem on the limit of authority flowing between nodes for personalized PageRank has been generalized. In this paper we want to extend those results to multimodal networks.
In particular we deal with a hypergraph type that may be used for describing multimodal network where a hyperlink connects nodes from each of the modalities.
We introduce a generalisation of PageRank for such graphs and define the respective random walk model that can be used for computations.
We state and prove theorems on the limit of outflow of authority for cases where individual modalities have identical and distinct damping factors.
\end{abstract}


\section{Introduction}

The notion of PageRank as a measure of importance of a Web page was introduced by Page \emph{et al.} in \citep{Page:1999}.
A formal introduction to PageRank can be found e.g.\  in \cite{Brezinski:2006}, or~\cite{WK11}.
The basic model was soon extended in diverse directions (like: personalized PageRank, topical PageRank, Ranking with Back-step, Query-Dependent PageRank, Lazy Walk Pagerank etc.) to support   numerous applications (Web page ranking, client and seller ranking, clustering, classification of Web pages, word sense disambiguation, spam detection, detection of dead pages etc.), as presented e.g.\  in a review   by  Langville and Meyer
\citep{LM06}, or by Berkhin \citep{Berkhin05asurvey}.

The success story of PageRank prompted many researchers to apply it also to bipartite graphs and other graphs representing multimodal relationship. Let us just mention studies by Link \citep{Link:2011}
concerning mutual evaluations of students and lecturers, by Deng et al., \citep{DLK09} on
reviewers and movies in a movie recommender systems, authors and papers in scientific literature, queries and URLs in query logs,
by Bauckhage  \citep{Bauckhage:2008} on performing image tagging or by Korner \cite{Korner:2010} on tagging in social networks or by Chen \cite{Chen:2015} on sentimental analysis.\footnote{Here we have the use-case of undirected graphs. As shown in \cite{Grolmusz:2012:undirectedPR}, the general intuition that PageRank for undirected graphs has a trivial form, is only valid under certain settings.}

For a number of reasons direct  transfer of PageRank to domains closely related to social networks has various deficiencies.
Therefore, a suitable generalization of PageRank to such structures is needed in order to retain both advantages of the multimodal graph representation and those of PageRank.

In a former paper \cite{Bipartite:2016}
a concept of Bipartite PageRank was developed and a theorem on limits of authority flow was proposed for it.

In the current paper we are interested in a further generalization to multimodal networks where we have more modalities.

The fundamental issues here seems to be the fact that unlike networks with one or two modalities, networks with more modalities are in fact hypergraphs, and not ordinary graphs. Hence a new concept of a random walk needs to be developed.

This paper is structured as follows:
Section \ref{sec:pagerank}  presents a brief introduction to PageRank idea.
In section \ref{sec-hg} we recall the concept of hypergraphs and then in section \ref{sec-related} we will review the literature on various ways of generalizing PageRank to hypergraphs.
Afterwards in section \ref{sec-multimodalnets}
we present our own generalisation of PageRank to  multimodal networks that we will call \emph{MuMoRank} and we present our major result on authority flow in such networks.
Section \ref{sec-conclusions} contains some final remarks.

\medskip
Our contributions are as follows:
\begin{itemize}
\item We propose a new ranking method for nodes in a multimodal hypergraph,
\item we investigate the flow limits for personalized version of this ranking
and prove respective theorems

\begin{itemize}
\item for damping factors identical for each modality and
\item for damping factors different for each modality
\end{itemize}
\end{itemize}

\section{Basic PageRank concepts}\label{sec:pagerank}

Usually, PageRank vector is introduced in a context of a random walk on a graph $\Gamma$ with edge weights $w_{i,j}$ for edges~$\{i,j\}$. The entries of the probability transition matrix~$\tilde{P}$, $\tilde{p}_{i,j} = w_{i,j}/d_j$ describe the probability of moving from page~$j$ to page~$i$. Here~$d_j$ stands for the degree of node~$j$. If~$\Gamma$ is undirected graph, $d_j = \sum_i w_{i,j}$, while if~$\Gamma$ is a directed graph then  more attention is required; see ~\cite{Page:1999} or~\cite[Ch. 4]{LM06} for details. The important thing, however, is that~$\tilde{P}$ is a column stochastic matrix, i.e. all entries are nonnegative and the sum of entries in each column is~$1$.  To approximate the behavior of a random surfer, Brin and Page invented so-called Google matrix~$P$, such that
\[P = (1-\zeta)\tilde{P} + \zeta\mathbf{s}\mathbf{1}^T.
\]
Here $0 < \zeta < 1$ is a jumping, or teleportation factor, ~$\mathbf{s}$ is a column-stochastic preference vector (i.e. all its elements are nonnegative and add up to one), and~$\mathbf{1}$ stands for the column vector of all ones. The role of the parameter~$\zeta$ is interpreted as follows: Being at a node~$j$, with the probability (1-$\zeta$) the surfer moves to one of the neighboring nodes, say~$i$, with the probability~$p_{i, j}$, and with the remaining probability $\zeta$ he gets bored (thus $\zeta$ is called also boring factor or authority emission rate) and hence jumps to any node in~$\Gamma$ with probability described by the preference vector~$\mathbf{s}$. In original formulation, $\mathbf{s}$ was the uniform distribution, but other choices are possible. Varying the entries of the preference vector leads to different forms of random walk personalization.
For instance, if these entries are proportional to the number of (in-going or out-going) edges, then we speak about  (authority-preferring or hub-preferring) preferential PageRank. In undirected networks, we speak only about preferential PageRank. But we can also treat the undirected links as pairs of links pointing in both directions.
In this paper, when talking about generalised graphs, we will always think about undirected, that is bidirectional links.
If, prior to any of the mentioned activities, the random surfer may decide with some probability, that he will stay at the node instead of walking or jumping, then we talk about lazy PageRank.

\medskip
The entire PageRank vector, denoted by~$\mathbf{r}$ is nothing but the stationary distribution of the Markov chain implied by the Google matrix~$P$. Formally,~$\mathbf{r}$ is the solution of the eigenvalue problem~$P\mathbf{r} = \mathbf{r}$, or more transparently
\[ \mathbf{r} = (1-\zeta)\tilde{P}\mathbf{r} + \zeta \mathbf{s}.
\]
Details concerning its computation can be found e.g.\  in  \cite{Chung14}, or \cite{Gleich15}.

The surfer model of PageRank may be equivalently expressed in terms of   ``authority flow''.
The $\mathbf{r}$ is understood as an authority distribution over the nodes of the network.
At discrete time points $t$, the authority $r_i$ of node $i$  flows via links to neighbouring nodes in the amount $r_i(1-\zeta)/deg(i)$ and the amount
$r_i\zeta$ flows to the so-called supernode, which redistributes immediately (at the same timepoint) authority among some or all nodes (nodes from the personalization set $U$), depending on the personalization model ($s$ vector).

Under stable conditions, authority $r_i$ of node $i$ remains the same at each time point, that is the amount of authority flowing in and flowing should be the same.
The presence of the supernode causes that the amount of authority flowing in and flowing out via ordinary links of a node is not the same.
So we have a net out-flow of authority $n_{a_U}$ from the set $U$ to its complement
via the ordinary links. This net flow must go back to $U$ via the supernode.
So $n_{a_U}=\sum_{k \not\in U}r_k\zeta$.

Both surfer interpretation and authority flow are of relevance in one-cluster clustering. A group of nodes $U$ is deemed to be a good cluster if a low amount of authority is flowing out of it or the surfer will stay longer in this set upon random walk.

\section{Hypergraphs}\label{sec-hg}

To extend the concept of PageRank
authority flow limitations, studied previously for unimodal \cite{Chen:2015} and bimodal networks \cite{Bipartite:2016},
to multimodal networks we have first to introduce the concept of random walk through such a network.\footnote{Our approach to defining hypergraphs and their PageRank differs from that used in e.g.\  by  Takai et al. \cite{Takai:2020} or Li and Milenkovic \cite{Li:2018:hglapl} as we need to address multimodality explicitly. }

Let us recall what the multiple modalities in a network may mean.
One such situation when multiple modalities are used is when a consumer buys a product and attaches a label ``good'' or ``bad'' to it. Another case is when an Internet   user evaluates a Web page and attaches one or more tags to it.
A teacher may evaluate a student and attach a label "passed exam" or "failed", etc. There are many other possibilities.
In the general case of a multimodal social network, a link ties together more than two objects
so that it is not an edge but rather a hyperedge, so we have to handle hypergraphs.

\medskip
Let us turn to a more formal description of the issue.
A hypergraph is generally defined as follows:
$$\HG=(\N, \HE)$$
where $\N$ is the set of nodes,
$\HE\subseteq 2^{\N}-\{\emptyset \}$ is the set of hyperedges, each hyperedge being a non-empty subset of $\N$.

\medskip
However, we are interested in a  special subtype of the hypergraphs, called M-uniform M-partite hypegraphs\footnote{ Such hypergraphs have been studied in different contexts, see e.g.\ \cite{Mycroft:2014}, \cite{Han:2020}, \cite{Schauz:2010}}.
Let $M$ be the number of distinct modalities and
$$\N=\N_1 \cup \N_2 \cup \dots \cup \N_M,$$ where $\N_i$ is a set of nodes of $i$-th   modality, and  $\N_j\cap \N_i = \{\emptyset\}$ for $i\ne j$, that is the intersection of modalities is empty.
Then a multimodal hypergraph can be defined as
$$\HGM=(\N,\HEM)$$
where  $card(\hem\cap \N_j)=1 $ for any $\hem \in \HEM$
 and $j=1,\dots,M$.
So each multimodal hyperedge is of the same cardinality ($M$).

\begin{example}\label{ex:leading} 
Let us illustrate the concepts just introduced with a small example.

\medskip
Assume a fictitious product evaluation database with three modalities:
\begin{itemize}
\itemsep=0.95pt
\item  users (	 		Eva, 	Mary, Bob, John,	Jane, Ann,	 Henry,	Max); the set of users shall be called $\N_u$,
\item  products (			TVset,
			VideoPlayer,
			Laptop,
			DVDPlayer,
			Smartphone,
			Netbook);  the set of products shall be called $\N_p$,
and \item  tags (handsome,
			welldesigned,
			beautiful,
			pretty,
			annoying,
			awful,
			worthless);  the set of tags shall be called $\N_t$.
\end{itemize}

Assume that the users have tagged the products as in Table \ref{tab:tagging}.

\begin{table}[ht!]
\vspace*{-4mm}
\caption{Product tagging example }\label{tab:tagging}
\begin{center}
\begin{tabular}{|r|lll|}
\hline
$\HEM$ element &User  name & Product name &  Tag\\
\hline
$\hem_1$&Eva & TVset & handsome\\
$\hem_2$&Eva & VideoPlayer & welldesigned\\
$\hem_3$&Eva & Laptop & awful\\
$\hem_4$&Eva & Netbook & awful\\
$\hem_5$&Mary & TVset & handsome\\
\hline
$\hem_6$&Mary & Smartphone & handsome\\
$\hem_7$&Mary & Laptop & beautiful\\
$\hem_8$&Mary & Netbook & beautiful\\
$\hem_9$&Bob & Laptop & beautiful\\
$\hem_{10}$&Bob & VideoPlayer & welldesigned\\
\hline
$\hem_{11}$&John & VideoPlayer & welldesigned\\
$\hem_{12}$&John & DVDPlayer & welldesigned\\
$\hem_{13}$&Jane & TVset & awful\\
$\hem_{14}$&Jane & VideoPlayer & beautiful\\
$\hem_{15}$&Jane & DVDPlayer & worthless\\
\hline
$\hem_{16}$&Jane & Smartphone & worthless\\
$\hem_{17}$&Ann & VideoPlayer & annoying\\
$\hem_{18}$&Ann & DVDPlayer & beautiful\\
$\hem_{19}$&Henry & Netbook & handsome\\
$\hem_{20}$&Henry & Laptop & awful\\
\hline
$\hem_{21}$&Henry & DVDPlayer & awful\\
$\hem_{22}$&Henry & Smartphone & awful\\
$\hem_{23}$&Max & Netbook & handsome\\
$\hem_{24}$&Max & Laptop & welldesigned\\
\hline
\end{tabular}

\end{center}
\end{table}

\medskip
A hypergraph corresponding to this product evaluation is constructed as follows. The set of nodes $\N$ will consist of users, products and tags: $\N=\N_u\cup\N_p\cup\N_t$. The set of hyperedges $\HEM\subset \N_u\times\N_p\times\N_t$ consists of $\hem_{1},\dots,\hem_{24}$ from Table \ref{tab:tagging}.
\end{example} 

\section{Ranks and random walks in multimodal networks}\label{sec-multimodalnets}

\subsection{Basic definitions and concepts}

Typically, PageRank is analyzed in the context of a random walk on the graph. A surfer moves through a graph from node to node via edges (links). If we want to deal with multimodality (with more than two modalities)  we encounter a problem, because the relations between objects (nodes) are no more represented by edges of a graph, but by hyperedges of a hypergaph.
A surfer cannot jump through a hyper edge to another object because there may be more than one ``at the other end'' to jump to.

So let us look at the multimodal network in a different way: let us define  ``generalized nodes'' to be either normal nodes (objects) or hyperedges (links). Define further the ``generalized edges'', each  linking a hyper edge with a node  that it is adjacent to.

\medskip
The ``generalized graph'' consisting of generalized nodes and generalized edges is an ordinary graph through which a surfer can go just like for traditional Pagerank. In this case we have a bipartite graph, with two kinds of nodes. There are however two issues to be still resolved:
\begin{itemize}
\item how to interpret  the jumps from the hyperedge back to the node that one was at before
\item how the flow of authority is to be organized for the jumps out of boredom.
\end{itemize}

The first issue is closely related to the concept of   so-called ``lazy walk'' that is that at a given moment one does or does not move or jump further (this may be thought of as simulation of a surfer staying for a shorter or a longer time at a node). This would require  to modify the theorem stated in \cite{Bipartite:2016} for lazy random walk.
The second issue is of course that one can get bored at different rates at different modalities.

\medskip
Define the generalized graph as follows:
$$\GG(\HGM)=(\GN, \GE)$$
where $\GN=\N\cup \HEM$, and $\GE\subseteq \N\times \HEM$
where $\Ge=(\n,\hem)$ iff $\n \in \hem$.

\begin{example}
    Let us continue the Example \ref{ex:leading}.
In this example $\GN=   \N_u\cup\N_p\cup\N_t\cup \{\hem_{1},\dots,\hem_{24}\}$.
$\GE$ includes bidirectional edges
$(\hem_1, Eva)$,  $(\hem_1, TVset)$, $(\hem_1, handsome)$,
$(\hem_2, Eva)$,   $(\hem_2,VideoPlayer)$,   $(\hem_2, welldesigned)$ and so on.
\end{example}

Obviously, the $\GG$ is an ordinary bipartite graph with two node subsets $\N$ and $\HEM$, but we cannot apply here directly the  approach from \cite{Bipartite:2016} to bipartite graphs.
This is because we do not want the hyper-edges $\HEM$ neither to emit nor to receive any authority to and from supernode.

Instead let us assume that each modality node subset $\N_i$ has its own supernode to which each node (of whatever modality)  emits authority and it redistributes the authority among its own nodes only. The modalities can have either a  common emission rate $\zeta$ for all of them or they can have separate emission rates $\zeta_m$ for each modality $m$.

Let us further assume that each modality has authorities that add up to one.

Let us consider a surfer through the graph $\GG$.
When in $\N_i$, the surfer may decide
either to perform a ``boring'' jump with probability $\zeta_m$ (jumping constant) to any modality and in that modality to any other node of the same modality while choosing the target node proportionally to its in/out degree and choosing it only if it belongs to the preferential set.
With probability $1-\zeta_m$ he chooses uniformly to go out through one of the edges leading to $\HEM$ set. When in $\HEM$, he may go out through any of the outgoing edges with uniform probability (landing in $\N$).

\begin{example} 
    Let us continue the
Example \ref{ex:leading}.
When the surfer is at the ``node'' $\hem_1$, he may jump either to $Eva$, or $TVset$, or $handsome$ with probability $1/3$ each. Being  at ``node'' $\hem_2$ he may jump either to  $Eva$ or to $VideoPlayer$ or to  $welldesigned $ also with probability 1/3.

When at ``node'' $Eva$, representing the modality ``users'', the jumping is more complicated. One can jump to one of the hyperedges $\hem_1, \hem_2, \hem_3,\hem_4$  with probability $\frac 14 (1-\zeta_u)$ or  one can choose a modality and jump to one of the nodes from the same modality with probability proportional to $\zeta_u$ times the degree of a given modality element. So $Eva$ has degree 4, $Mary$ 4, $Bob$ 2 and so on.
\end{example}

For each modality and for each node we define \emph{MuMoRank} as the probability that the surfer mentioned above finds himself in this node given he is in this modality.

Denote by $\mathbf{r_\N}$ a (column) vector of \emph{MuMoRanks}: $r_{\N,j}$ will mean the \emph{MuMoRank} of node $j$. All elements of $\mathbf{r_\N}$ are non-negative and their sum equals 1 for each subset $\N_i$ of $\N$ (that is modality $i$). With $\mathbf{r_{\HEM}}$ we will denote a (column) vector of supplementary ranks of hyperedges (\emph{Hyperedge-MuMoRanks}).

\medskip
Let
$$\mathbf{P_{\N\rightarrow \HEM}} = [p_{\N\rightarrow \HEM,i,j}]$$ be a matrix such that if there is a link from node $j$ to hyperedge $i$, then $p_{\N\rightarrow \HEM,i,j}=\frac{1}{deg(j)}$,
where $deg(j)$ is the  degree of node $j$.
In other words, $\mathbf{P_{\N\rightarrow \HEM}}$ is column-stochastic matrix satisfying $$\sum_i p_{\N\rightarrow \HEM,i,j} = 1$$ for each column $j$. If a node has a degree equal to 0, then prior to construction of $\mathbf{P_{\N\rightarrow \HEM}}$ this node is removed from the network.

\medskip
Further, let $$\mathbf{P_{\HEM\rightarrow \N}} = [p_{\HEM\rightarrow \N,i,j}]$$ be a matrix such that if there is a link from hyperedge $j$ to node $i$, then $p_{\HEM\rightarrow \N,i,j}=\frac{1}{deg(j)}$,
where $deg(j)$ is the degree of hyperedge $j$.
In other words, $\mathbf{P_{\HEM\rightarrow \N}}$ is column-stochastic matrix satisfying $$\sum_i p_{\HEM\rightarrow \N,i,j} = 1$$ for each column $j$. There exists no hyperedge  with  degree equal to 0.
In fact each has the degree equal to number of modalities.

\medskip
Under these circumstances we have
\begin{equation} \label{eq:rdef}
\begin{split}
\mathbf{r}_{\HEM} &=(\mathbf{I}-\Zeta){\cdot} \mathbf{P_{\N\rightarrow \HEM}}{\cdot}\mathbf{r_\N} \\
\mathbf{r}_{\N} &= \mathbf{P_{\HEM\rightarrow \N}}{\cdot}\mathbf{r_{\HEM}}
+ \Zeta {\cdot} \mathbf{s}
\end{split}
\end{equation}

\noindent where $\mathbf{s}$ is the so-called ``initial'' probability distribution (i.e. a column vector with non-negative elements summing up to 1 for each modality separately) that is also interpreted as a vector of node preferences, and $\Zeta$ is the diagonal matrix of damping factor values corresponding to appropriate modalities~$m$.

The limit theorem for modalities is similar to
the lazy walk preferential
\cite[Theorem 2]{Bipartite:2016}.

Let us consider a generalization of personalized PageRank to hyper{}graphs, in the spirit just mentioned for bipartite networks. Let $U^{(i)}$ be the set of nodes of modality $i$ to which one jumps preferentially upon being bored, and let such a set exist for each modality.
This means that $s_j=\frac{1}{|U^{(i)}|}$ if node~$j$ belongs to modality $i$ and lies in the set $U^{(i)}$, and is zero otherwise.

\begin{example}
Let us continue Example \ref{ex:leading}.

\medskip
Assume that the damping factors $\zeta$ are equal to 0.3 for users, 0.2 for products and 0.1 for tags.
Assume also that our set of preferred nodes consists of: \smallskip\\
				\{Eva, Mary, Henry,  beautiful, awful, Laptop, Netbook \}. \smallskip\\
Hub-preferring walk is assumed. 		

\medskip
In Table \ref{tab:MuMoRank} we have the resulting \emph{MuMoRanks}.

\begin{table}[ht!]
\caption{\emph{MuMoRanks} of nodes in the hypergraph derived from Table \ref{tab:tagging}}\label{tab:MuMoRank}
\begin{center}
\begin{tabular}{|ll|}
\hline
Node name &  {MuMoRank} \\
\hline
 Eva	  &  0.222723\Bem{7898750969}	\\
 Mary	  &  0.227777\Bem{17270236}	\\
 Bob	  &  0.061828\Bem{005075369515}	\\
 John	  &  0.033909\Bem{153659620814}	\\
 Jane	  &  0.100468\Bem{20687444284}	\\
 Ann	  &  0.045146\Bem{4448214134	}\\
 Henry	  &  0.239510\Bem{27791757953}	\\
 Max	  &  0.068636\Bem{94887041327}	\\
\hline
 TVset	  &  0.097783\Bem{4762379729}	\\
 VideoPlayer	 & 0.105357\Bem{9150501943}	\\
 Laptop	 & 0.33408509\Bem{623747196}	\\
 \hline
\end{tabular}
\begin{tabular}{|ll|}
\hline
Node name &  {MuMoRank} \\
\hline
 DVDPlayer	  &  0.10552\Bem{136952069643}	\\
 Smartphone	  &  0.09269\Bem{5605367122}	\\
 Netbook	  &  0.26455\Bem{65373828387	}\\
 \hline
 handsome	  &  0.17491\Bem{834988889507}	\\
 welldesigned	  &  0.11119\Bem{309198650744}	\\
 beautiful	  &  0.28821\Bem{5407332984}	\\
 pretty	  &  0.0	\\
 annoying	& 0.01555\Bem{1677185920565}	\\
 awful	 & 	  0.37155\Bem{624749822336}	\\
 worthless	  &  0.03856\Bem{522590376586}	\\
\hline
\end{tabular}

\end{center}
\end{table}

Within each modality the sum of \emph{MuMoRanks} sums up to 1.
\end{example}

In Equations (\ref{eq:rdef}), the steady state of rank distribution was assumed. But for purposes of computation of authority flow, we will consider the ranks on their way to stable conditions in time steps $t=0,1,\dots$.
\begin{equation} \label{eq:rdeftime}
\begin{split}
\mathbf{r}_{\HEM}[t+1] &=(\mathbf{I}-\Zeta){\cdot} \mathbf{P_{\N\rightarrow \HEM}}{\cdot}\mathbf{r_\N}[t] \\
\mathbf{r}_{\N}[t+1] &= \mathbf{P_{\HEM\rightarrow \N}}{\cdot}\mathbf{r_{\HEM}} [t]
+ \Zeta {\cdot} \mathbf{s}
\end{split}
\end{equation}

The next subsections will deal with estimation of limits of
authority flow within a modality hypergraph. The estimation is based on structural features of the hypergraph.

\subsection{Damping factors equal zero}

Assume  that at the moment $t$ we have the following state of authority distribution: node $j$ of modality $i$
contains $$r_{j}[t]=\frac{deg(j)}{\sum_{k\in \N_i} deg(k)}.$$

Assume that $\zeta=0$ and consider the moment $t{+}1$.
Node $j$ of the modality $i$ passes
into each outgoing link the authority (to each incident hyperedge)
$ \frac{1}{\sum_{k\in \N_i} deg(k)} $.
 This quantity is obviously identical for each node  of the modality $i$. But as each hyperedge is connected to exactly one node of each modality, hence $\sum_{k\in \N_i} deg(k)$ is also identical for each modality, and in effect each node gives each hyperedge the same quantity of authority. As the hyperedge distributes the authority evenly, the same amount returns to each node again. So we have equilibrium - the authority distribution does not change.

\subsection{Damping factors identical for all modalities}

Let us turn to the case $\zeta_{i}$'s being greater than zero.   We will elaborate a couple of theorems limiting the flow of authority in the hypergraphs.
Consider a personalization set $U=U^{(1)}\cup \dots \cup U^{(M)}$, where
$U^{(1)}\subseteq \N_1,\dots, U^{(M)}\subseteq \N_M.$

Let us make first the simplification that all the $\zeta_{i}$'s \emph{are the same} -- just equal  $\zeta$.
The amount of authority passed to a node consists of two parts: a variable one being a share of the authority at the feeding end of the link and a fixed one coming from a supernode. So, by increasing the variable part say in a vicinity of  a node we come to the point that the receiving end gets less authority than was there on the other end of the link because of the ``redistribution'' role of the supernode(s).

\medskip
Let us  look for the upper bound of authority $d$
that for any plain node $k$ of modality $i$: $r_{k}[t]\le d\cdot deg(k)$ at any point of time $t$.
 A plain node $k$ has the authority $r_{k}[t]$ at the point of time $t$. $r_{i}[t]\cdot (1-\zeta)\le d\cdot (1-\zeta)\cdot deg(k)$ flows to the hyperedges     and $r_{k}[t]\cdot \zeta$ flows to the node's modality supernode.
The supernode gets from all modality nodes together $\zeta$ authority.
And it redistributes it to the modality nodes proportionally to node degrees, so that node $k$ gets $\frac{\zeta\cdot deg(k) }{\sum_{v \in U^{(i)}}deg(v)}$ if $k$ belongs to $U^{(i)}$, and zero otherwise.
On the other hand, the hyperedge obtains at most $d\cdot(1-\zeta)\cdot M$ authority from all plain nodes. And it redistributes it back to its $M$ intersecting plain nodes so that each gets at most $d\cdot(1-\zeta)$. Therefore, if $d$ should be the upper limit of plain node authority divided by the node degree, then:
$$ d{\cdot}(1-\zeta ) +
 \frac{\zeta }{\sum_{v \in U^{(i)}}deg(v)}  \le d.$$
Hence, for each modality $i$
$$d\ge \frac{1}{\sum_{v\in U^{(i)}} deg(v)} .$$
So we obtain a satisfactory $d$ when
$$d_{sat}=
\max_{i=1,\dots,M}\frac{1}{\sum_{v\in U^{(i)}} deg(v)}$$

Let us formulate a theorem for \emph{MuMoRanks} (multimodal  PageRank) limiting the outflow of authority analogous to the classical
\cite[Theorem 2]{Bipartite:2016}.
We would guess that the chance of a surfer/authority staying within a subset of nodes $U$ would depend on the internal link structure of the network. The theorem below shows, however, that we can find an upper bound on authority outflow with a very limited knowledge of the internal structure -- just the sum of node degrees within each modality is a sufficient statistics.

\begin{theorem} \label{thMPRlimit}
For the preferential personalized \emph{MuMoRank}  we have
$$
n_{a_U}
\le (1-\zeta ) \frac{|\partial{U}|} {\displaystyle \min_{i=1,\dots ,M}(HVol(U^{(i)}))} \vspace*{-2mm}$$
where
\begin{itemize}
\itemsep=0.95pt
\item $n_{a_U}$ is the net authority out-flow from the  set $U$ into $\N / U$.
\item $|\partial {U} |=\sum_{\hem \in \HEM}\frac{l_{n,\hem} \cdot  l_{o,\hem}}{M}$ is the
sum over all hyperedges of $\frac{l_{n,\hem} \cdot  l_{o,\hem}}{M}$,
where $l_{n,\hem}$ is the number of links from $U$ intersecting with the hyperedge $\hem$,
 $l_{o,\hem}$ is the number of links from not $U$ intersecting with the hyperedge $\hem$, (note that $l_{n,\hem}+l_{o,\hem}=M$.)
\item  $HVol(U^{(i)}) =\sum_{v\in U^{(i)}} deg(v)$ is the sum of  degrees of nodes in  $U^{(i)}$ (number of   hyperedges intersecting with  $U^{(i)}$ -  capacity of $U^{(i)}$) \QED
\end{itemize}
\end{theorem}

\begin{proof}
Recall that, under stable conditions, the outflow of authority from $U$ into the rest of ordinary nodes (that is the amount of authority flowing out but not being compensated by authority flowing in) has to be compensated by the flow from outside via the supernodes. So instead of asking about authority outflow, we can ask:
``How much authority from outside of $U$ can flow into $U$ via supernodes at the point of stability?''. Each node $k$ outside of $U$ gives $\zeta$ share of its authority to the supernodes (not getting anything from them).
Let  $p_{i,o}=\sum_{k \in \N_i \backslash U^{(i)}}r_k$ be the sum of authorities from the set $\N_i \backslash U^{(i)}$.
Then $n_{a_U}=\sum_{i=1}^M p_{i,o}\zeta$.

\medskip
The proof  is analogous to the case of classical PageRank presented in \cite{Bipartite:2016}, using now the quantity $d_{sat}$ we have just introduced.
The idea is that the authority outflowing through outlinks from $U$ to the remaining nodes  must enter again the set $U$ via
the respective supernodes.
So the quantity $\sum_{i=1}^M  p_{i,o}\zeta$ is the authority   re-entering  the set $U$ via
the respective supernodes.
On the other hand $(1-\zeta)d_{sat}$  is what at most leaves the set $U$ via a link leading outside of $U$. $l_{n,\hem}$ links lead from $U$ to the hyperedge $\hem$, this amount is divided by $M$ and sent back to the $M$ intersecting nodes, out of which $l_{o,\hem}$ lead the authority outside of $U$.
So the outflow via this hyperedge is at most  $\frac{l_{n,\hem} \cdot  l_{o,\hem}}{M}\cdot (1-\zeta)d_{sat}$. If we sum this last expression up over all hyperedges $\hem \in \HEM$, we get:
\begin{eqnarray*}
& \displaystyle (1-\zeta)d_{sat}|\partial{U}|=
(1-\zeta)\max_{i=1,\dots ,M}\frac{1}{\sum_{v\in U^{(i)}} deg(v)}|\partial{U}|= &
\\[4pt]
& \displaystyle (1-\zeta)
\frac{|\partial{U}|}{\min_{i=1,\dots ,M}\sum_{v\in U^{(i)}} deg(v)}&
\end{eqnarray*}

This completes the proof.
\end{proof}
\eject

We can refine this reasoning assuming that we are not looking
for a single $d$ but for a set of separate $d$'s for each modality.

\begin{theorem} \label{thMPRlimitd0}
For the preferential personalized \emph{MuMoRank}
we have
$$n_{a_U}\le
  (1-\zeta ) \Bigg(\Big( \sum_{\hem\in \HEM }\frac{l_{o,\hem}}{M}\sum_{i:nd(i,\hem)\in U}\frac{\zeta }{HVol(U^{(i)})}\Big)+
  \Big(\frac1M\sum_{i=1}^M\frac{(1-\zeta) }{ HVol(U^{(i)})}   \sum_{\hem\in \HEM }\frac{l_{o,\hem}\cdot l_{n,\hem}}{M}  \Big)\Bigg)
$$
where
\begin{itemize}
\itemsep=0.95pt
\item $n_{a_U}$ is the net authority out-flow from the  set $U$ into $\N / U$.
\item  $nd(i,\hem)$ denotes the node of modality $i$ intersecting with the hyperedge $\hem$.
\item  $l_{n,\hem}$ is the number of links from $U$ intersecting with the hyperedge $\hem$,
\item
 $l_{o,\hem}$ is the number of links from not $U$ intersecting with the hyperedge $\hem$, (note that $l_{n,\hem}+l_{o,\hem}=M$.)
\item  $HVol(U^{(i)})$ is the sum of  degrees of nodes in  $U^{(i)}$ (number of   hyperedges intersecting with  $U^{(i)}$ -  capacity of $U^{(i)}$) \QED
\end{itemize}
\end{theorem}

\begin{proof}
Let  $p_{i,o}=\sum_{k \in \N_i \backslash U^{(i)}}r_k$ be the sum of authorities from the set $\N_i \backslash U^{(i)}$,
Then $n_{a_U}=\sum_{i=1}^M p_{i,o}\zeta$.
Consider   the situation that on the one hand if a node of $i$th modality has at most $d^{(i)}$ amount of authority per link, then it sends to a hyperedge at most $d^{(i)}{\cdot}(1-\zeta)$ authority via the link.
The receiving hyperedge redistributes
to each of  the  links at most $$\frac1M\sum_id^{(i)}{\cdot}(1-\zeta)$$
Any node belonging to an $U^{(i)}$  gets additionally from the supernode exactly
$$\frac{\zeta }{\sum_{v \in U^{(i)}}deg(v)}$$  authority per its link.
We seek   $d^{(i)}$ such that these two components do not exceed $d^{(i)}$ together, i.e.
$$ \frac1M\sum_id^{(i)}{\cdot}(1-\zeta)
 +
 \frac{\zeta }{\sum_{v \in U^{(i)}}deg(v)}  \le d^{(i)} $$

 Based on this formulation we can start to seek for   $d$'s of the form
$$d^{(i)}=d_0+ \frac{\zeta }{\sum_{v \in U^{(i)}}deg(v)}  $$
where $d_0$ is some base authority.\footnote{Base authority should be understood as a non-negative number expressing the minimum amount of authority per link of a node.}

\eject

This leads immediately to
$$ d_0\cdot (1-\zeta)+ \frac1M\sum_i\frac{\zeta }{\sum_{v \in U^{(i)}}deg(v)}{\cdot}(1-\zeta)
 +
 \frac{\zeta }{\sum_{v \in U^{(i)}}deg(v)}  \le
d_0+ \frac{\zeta }{\sum_{v \in U^{(i)}}deg(v)}
 $$
$$   \frac1M\sum_i\frac{\zeta }{\sum_{v \in U^{(i)}}deg(v)}{\cdot}(1-\zeta)
   \le d_0 \zeta
 $$

So that the satisfactory $d_0$ would be
$$ d_{0,sat}=  \frac1M\sum_i\frac{1 }{\sum_{v \in U^{(i)}}deg(v)}{\cdot}(1-\zeta)
 $$
$U$ is left via intersections with $\hem$ by the amount of authority equal to
$\sum_{i, nd(i,\hem)\in U} (1-\zeta)d^{(i)} $.
At $\hem$,  this amount is divided by $M$ and sent back to the $M$ intersecting nodes, out of which $l_{o,\hem}$ lead the authority outside of $U$.  So the outflow via this hyperedge is at most
$\frac{ l_{o,\hem}}{M}\sum_{i: nd(i,\hem)\in U} (1-\zeta)d^{(i)} $.
If we sum this last expression up over all hyperedges $\hem \in \HEM$, we get the outflow of authority of at most:
\begin{equation*}
\sum_{\hem\in\HEM}\frac{ l_{o,\hem}}{M}\sum_{i: nd(i,\hem)\in U} (1-\zeta)d^{(i)}
\end{equation*}
\begin{equation*}
=(1-\zeta)\sum_{\hem\in\HEM}\frac{ l_{o,\hem}}{M}\sum_{i: nd(i,\hem)\in U} \left(d_0+ \frac{\zeta }{\sum_{v \in U^{(i)}}deg(v)} \right)
\end{equation*}
\Bem{
\begin{equation*}
=(1-\zeta)
\left(
\left(d_0\sum_{\hem\in\HEM}\frac{ l_{o,\hem}}{M}\sum_{i: nd(i,\hem)\in U}1 \right)
+
\left(
\sum_{\hem\in\HEM}\frac{ l_{o,\hem}}{M}\sum_{i: nd(i,\hem)\in U}
\frac{\zeta }{\sum_{v \in U^{(i)}}deg(v)} \right)
\right)
\end{equation*}
}
\begin{equation*}
=(1-\zeta)
\left(
\left(d_0\sum_{\hem\in\HEM}\frac{ l_{o,\hem}\cdot l_{n,\hem}}{M} \right)
+
\left(
\sum_{\hem\in\HEM}\frac{ l_{o,\hem}}{M}\sum_{i: nd(i,\hem)\in U}
\frac{\zeta }{\sum_{v \in U^{(i)}}deg(v)} \right)
\right)
\end{equation*}
This implies the theorem.
\end{proof}

\subsection{Damping factors different between modalities}

We can repeat these considerations assuming that the damping factors $\zeta$ \emph{can differ between the modalities}.
However, when dealing with equal $\zeta$s for each modality, we were allowed to assume that supernodes do not intersect. Now we have to admit that supernodes exchange in equal shares their authority with other supernodes. This mimics the assumption that a surfer, bored with ordinary walk, may select (uniformly) a modality and then jump to nodes in that modality with adequate link-based proportionality.

\begin{theorem} \label{thMPRlimituneqzeta}
For the preferential personalized \emph{MuMoRank}
we have
$$ n_{a_U}
\le  \frac{|\partial{U^\zeta}| }{\min_{i=1,...,M}(HVol(U^{(i)}))}$$
\eject
where
\begin{itemize}
\itemsep=0.95pt
\item $n_{a_U}$ is the net authority out-flow from the  set $U$ into $\N / U$.
\item  $nd(i,\hem)$ denotes the node of modality $i$ intersecting with the hyperedge $\hem$.
\item $|\partial {U^\zeta} |
=\sum_{\hem\in\HEM}
\frac{l_{o,\hem} \cdot  \sum_{i: nd(i,\hem)\in U} (1-\zeta^{(i)})}{M}
$ is the
sum over all hyperedges of $\frac{l_o \cdot  \sum_{j\in J} (1-\zeta^{(i)})}{M}$,
where
 $l_{o,\hem}$ is the number of links from not $U$ intersecting with this hyperedge $\hem$,
\item  $HVol(U^{(i)})$ is the sum of  degrees of nodes in  $U^{(i)}$ (number of   hyperedges intersecting with  $U^{(i)}$ -  capacity of $U^{(i)}$)\QED
\end{itemize}
\end{theorem}

\begin{proof}
Let  $p_{i,o}=\sum_{k \in \N_i \backslash U^{(i)}}r_k$ be the sum of authorities from the set $\N_i \backslash U^{(i)}$,
Then $n_{a_U}=\sum_{i=1}^M p_{i,o}\zeta^{(i)}$.
Let us consider nonzero   $\zeta^{(i)}$s for modalities i=1,...,M.
Let  $p_{i,o}=\sum_{k \in \N_i \backslash U^{(i)}}r_k$ be the sum of authorities from the set $\N_i \backslash U^{(i)}$,
Then $n_{a_U}=\sum_{i=1}^M p_{i,o}\zeta$.

\medskip
Let us seek the upper bounding amount of authority $d$
that for any plain node $k$ of modality $i$: $r_{k}[t]\le d\cdot deg(k)$ at any point of time $t$.

 A plain node $k$ has the authority $r_{k}[t]$ at the point of time $t$. $r_{i}[t]\cdot (1-\zeta^{(i)} )\le d\cdot (1-\zeta^{(i)} )\cdot deg(k)$ flows to the hyperedges and $r_{k}[t]\cdot \zeta^{(i)} $ flows to the node's modality supernode.

The supernode gets from all modality nodes together $\zeta^{(i)} $ authority.
It distributes this authority among all the modalities so that each modality supernode gets from modality $i$ amount $\frac{\zeta^{(i)}}{M}$ which in turn redistribute it among their modality nodes.
So the node $k$ gets from its supernode
$ \frac{
\deg(k)
   \frac1M
   \sum_i\zeta^{(i)} }
   {
\sum_{v \in U^{(i)}}deg(v)} $
 if $k$ belongs to $U^{(i)}$, and zero otherwise.  On the other hand, the hyperedge obtains at most $\sum_{i=1}^M d\cdot(1-\zeta^{(i)} )$ authority from all plain nodes. And it redistributes it back to its $M$ intersecting plain nodes so that each gets at most
 $\frac{1}{M}\sum_{i=1}^M d\cdot(1-\zeta^{(i)} )$.
 Therefore, if $d$ should be the upper limit of plain node authority divided by the node degree,
and if   we seek   a single $d$ for all modalities, then we would proceed as follows.

Let us seek the amount of authority $d$ such that multiplied by the number of  links of a sending node will be not lower than the authority of this node and that after the time step
its receiving node would have also amount of authority equal or lower than $d$ multiplied by the number of its in-links.

\medskip
That is we want to have that:
$$ d{\cdot}(1-\zeta^{(i)} ) +
 \frac{\frac1M\sum_i\zeta^{(i)} }{\sum_{v \in U^{(i)}}deg(v)}  \le d $$

This implies immediately, that %
$$
 \frac{\frac1M\sum_i\zeta^{(i)} }{\sum_{v \in U^{(i)}}deg(v)}  \le d \zeta^{(i)}$$
$$  d\ge
 \frac{\frac1M\sum_i\zeta^{(i)} }{\zeta^{(i)}\sum_{v \in U^{(i)}}deg(v)}   $$
 for each modality $i$.

\eject
Hence we get  a satisfactory $d$ when
$$d_{sat}=
\max_{i=1,...,M}
 \frac{\frac1M\sum_i\zeta^{(i)} }{\zeta^{(i)}\sum_{v \in U^{(i)}}deg(v)}   $$
By reasoning as in the proof of Theorem \ref{thMPRlimitd0} about the authority flown into and out of a hyperedge, we get the inequality of this theorem.
 \end{proof}

Alternatively think of separate $d$s for all the modalities.
\begin{theorem} \label{thMPRlimitd0uneqzeta}
For the preferential personalized \emph{MuMoRank}
we have
$$  n_{a_U}
\le
 \sum_{\hem\in \HEM } \frac{l_{o,\hem}}{M}\sum_{i:nd(i,\hem)\in U} (1-\zeta^{(i)}) \left(  \frac{\overline{\zeta}  }{HVol(U^{(i)}) }+ \frac1M\sum_i\frac{ 1-\zeta^{(i)}    }{HVol(U^{(i)})}\right)
$$
where
\begin{itemize}
\itemsep=0.95pt
\item $n_{a_U}$ is the net authority out-flow from the  set $U$ into $\N / U$.
\item
 $l_{o,\hem}$ is the number of links from not $U$ intersecting with this hyperedge $\hem$,
\item  $HVol(U^{(i)})$ is the sum of  degrees of nodes in  $U^{(i)}$ (number of   hyperedges intersecting with  $U^{(i)}$ -  capacity of $U^{(i)}$),
\item
$\overline{\zeta}=\frac1M\sum_i \zeta^{(i)}$ is an ``average'' $\zeta$. \QED
\end{itemize}
\end{theorem}


\begin{proof}
Let  $p_{i,o}=\sum_{k \in \N_i \backslash U^{(i)}}r_k$ be the sum of authorities from the set $\N_i \backslash U^{(i)}$,
Then $n_{a_U}=\sum_{i=1}^M p_{i,o}\zeta^{(i)}$.
Consider   the situation
that
on the one hand if a node of $i$th modality   has at most $d^{(i)}$ amount of authority per link, then it sends to a hyperedge
at most $$d^{(i)}{\cdot}(1-\zeta^{(i)})$$ authority via the link.
The receiving hyperedge redistributes
to each of  the  links at most $$\frac1M\sum_id^{(i)}{\cdot}(1-\zeta^{(i)})$$
Any node belonging to an $U^{(i)}$  gets additionally from the supernode exactly
$$\frac{\overline{\zeta} }{\sum_{v \in U^{(i)}}deg(v)}$$
with $\overline{\zeta}=\frac1M\sum_i \zeta^{(i)}$
authority per its link.
We seek   $d^{(i)}$ such that these two components do not exceed $d^{(i)}$ together.

So
$$ \frac1M\sum_id^{(i)}{\cdot}(1-\zeta^{(i)})
 +
 \frac{\overline{\zeta}  }{\sum_{v \in U^{(i)}}deg(v)}  \le d^{(i)} $$

Based on this formulation we can start to seek for such $d$s that
$d^{(i)}=d_0+ \frac{\overline{\zeta}  }{\sum_{v \in U^{(i)}}deg(v)}  $
where $d_0$ is some base authority.\footnote{Again, base authority should be understood as a non-negative number expressing the minimum amount of authority per link of a node.}
This leads immediately to
$$ d_0\cdot \frac1M\sum_i(1-\zeta^{(i)})+ \frac1M\sum_i\frac{\overline{\zeta}  }{\sum_{v \in U^{(i)}}deg(v)}{\cdot}(1-\zeta^{(i)})
 +
 \frac{\overline{\zeta} }{\sum_{v \in U^{(i)}}deg(v)}  \le
d_0+ \frac{\overline{\zeta}  }{\sum_{v \in U^{(i)}}deg(v)}
 $$
$$ d_0\cdot  (1-\overline{\zeta} )+ \frac1M\sum_i\frac{\overline{\zeta}  }{\sum_{v \in U^{(i)}}deg(v)}{\cdot}(1-\zeta^{(i)})
   \le
d_0
 $$
$$   \frac1M\sum_i\frac{\overline{\zeta}  }{\sum_{v \in U^{(i)}}deg(v)}{\cdot}(1-\zeta^{(i)})
   \le
d_0 \overline{\zeta}
 $$
$$   \frac1M\sum_i\frac{ 1-\zeta^{(i)}    }{\sum_{v \in U^{(i)}}deg(v)}
   \le
d_0
 $$

From which the   satisfactory $d_0$ can be derived
like
$$d_{0,sat}=   \frac1M\sum_i\frac{ 1-\zeta^{(i)}    }{\sum_{v \in U^{(i)}}deg(v)}
 $$

By reasoning as in the proof of Theorem \ref{thMPRlimitd0} about the authority flown into and out of a hyperedge, we get the inequility of this theorem.
\end{proof}

\begin{example}
Let us continue Example \ref{ex:leading}
The observed outflow of authority from our preferred nodes to the other amounts to
 0.2072\Bem{91135522084}.
Note that 		 $HVol(U^{users})= 12$,
		  $HVol(U^{products})= 9$,
		  $HVol(U^{tags})= 11$.
We get then $d_{sat}=0.1818\Bem{181818181818}$,
		$|\partial{U^\zeta}| =6.8666\Bem{66666666666}$.
Therefore according to Theorem \ref{thMPRlimituneqzeta} the upper authority outflow limit amounts to
 $0.7629\Bem{62962962963}$ (which is much higher than the actual one, this is a typical issue with small networks).

\medskip
Alternatively  if we use Theorem \ref{thMPRlimitd0uneqzeta} to find the bounds on authority, we get the following estimates.
$ d_{0,sat}=0.0763\Bem{4680134680134}$,
		$d_{sat}^{users}=0.0930\Bem{13468013468}$,
		$d_{sat}^{products}=0.0985\Bem{6902356902356}$,
		$d_{sat}^{ tags}=0.0945\Bem{2861952861952}$.
and then we get a lower bound of  0.6516 on authority outflow, which is slightly better.
\end{example}

One topic was not touched above, namely that of convergence.
But the convergence can be looked for in an analogous way as done for the HITS (consult e.g.\  \citep[Ch. 11]{LM06}).

\section{Discussion -- various ways of generalizing PageRank to hypergraphs}\label{sec-related}

One stream of research is not related to multimodal generalizations, but rather is concentrated on stabilizing or accelerating PageRank computations.
This idea is represented e.g.\  by  Petrovic \cite{Hyperpagerank} and
Berlt et al. \cite{Berlt:2010}.
Essentially instead of single pages, disjoint groups of them
(like domains or hosts) are considered
and with each link from a page A to page B a hyperedge is associated
containing the node B and all the nodes from the group to which A belongs.
Such an approach is claimed to contribute to fighting spam.
In the similar stream, Bradley \cite{Bradley:2005hypergraphpartitioning}
uses the concept of hypergraph partitioning for assignment of tasks for efficient PageRank computation by available processors.

Hotho et al. \cite{Hotho:2006,Hotho:2006folkrank} deal with generalisation of PageRank to folksonomies, that is hypergraphs with three modalities.
They convert the hypergraph to an ordinary undirected graph,
with the set of nodes being the union of all three modalities and edges
occurring between nodes that belong to some common hyperedge.
Then they apply the lazy version of traditional PageRank.
Bellaachia \cite{Bellaachia:2013}
defines, following the mentioned and other approaches,  a random walk over the hypergraph in such a way as if there were two kinds of nodes,
ordinary nodes and hyperedges which are connected by ordinary (undirected) edges whenever a
hyperedge is incidental with a node.
In such a graph traditional PageRank is applied.
A generalisation to weighted and personalised graphs is done in a straight-forward way.
Cooper \cite{Cooper:2013}
considers random walks over hypergraphs in the same spirit
concentrating on regular forms of the hypergraph (with a fixed size hypergraphs or nodes incident with a fixed number of hyperedges).
the goal is to compute the cover time.  Neubauer
\cite{Neubauer:2009} seeks communities in   hypergraphs via random walks.

Our approach to generalisation of PageRank to hypergraphs differs from the ones mentioned above in a number of ways.
Unlike Cooper  \cite{Cooper:2013}, we do not constrain ourselves to regular graphs.
Unlike Hotho et al.  \cite{Hotho:2006,Hotho:2006folkrank} we do not restrict our approach to only three modalities.
Note also that in case of three modalities only, Bellaachia \cite{Bellaachia:2013}
approach to PageRank computation is essentially equivalent to that of
\cite{Hotho:2006,Hotho:2006folkrank}.
The essential difference to our approach is the following:
We consider all the modalities separately.
We think that it is not a usable information whether or not a user has a higher rank than a Web page, or a client than a product. We would rather compare members of modalities.
As we handle the modalities separately,   each can have a separate decay factor which is not the case in the abovementioned approaches. Hence the values of ranks will be different. The differences will become more visible, if the number of nodes being "personalised" differs between the modalities.
Last not the least we introduce the concept of separate supernodes for each of the modalities so that the authority is not lost by one modality and passed to another because we consider the modalities as separate, incomparable concepts.
This separation is an important   difference between our approach and the existent ones. And we consider such an assumption as a quite natural one.

It is worth noting that there exist notions of "HyperRank" not related to PageRank.
Xu \cite{Xu:2012} uses hypergraphs to associate images, their tags and geolocations.
It creates its own "HyperRank" not being a generalisation of PageRank, but rather a ranking of objects in response to a user query, with an explicit ranking formula. But
\cite{Bu:2010} applies the idea of hypergraph in music recommendation.
As previously, the ranking induced has not been derived from PageRank, but is rather a closed-form approach.

\eject
Li and Milenkovic \cite{Li:2018:hglapl} introduced the concept of personalized Pagerank for hypergraph relying on introduction of a Laplacian for hypergraphs. Their goal is to cluster hypergraphs. They abstract from the notion of random walk, hoewever, they introduce weighted hyperedges.

Schaub et al. \cite{Schaub:2018:simplicialComplexes} propose generalization of PageRank to simplicial complexes which are special cases of hypergraphs, (V,E),  with "complete" hyperedges that is $\forall_{e\in E}$ if $ x \subseteq e $ then $x \in E$, and apply such hypergraphs to analyse
trajectory data of ocean drifters near Madagascar and to investigate  book co-purchasing.

\medskip
Apparently in none of these approaches the problem that we want to deal with here was considered:
\begin{itemize}
\itemsep=0.95pt
\item sealed off modalities from which authority does not flow out
\item varying damping  factor (jumping constant) in various modalities
\item the problem of outflow of authority from a selected set of nodes
\end{itemize}

\section{Concluding remarks}\label{sec-conclusions}

In this paper we proposed   a hypergraph type usable for describing multimodal network where a hyperlink connects nodes from each of the modalities, like users, tags, products etc..
We have introduced a novel approach to the concept of
multimodal PageRank for such a network, ranking each modality separately, because we consider the modalities as incomparable in their rankings.
We  defined the respective random walk model that can be used for computations.

We have   proposed (upper) limits for
the flow of  authority in a multimodal hypergraph
and  proved theorems on the limit of outflow of authority for cases where individual modalities have identical and distinct damping factors.
Finally we  illustrated the concepts with an example.

These limits can be used
in a number of ways, including verification of validity of   clusters in hyper graphs. It is  quite a common assumption that the better the cluster the less authority flows out of it.
The theorems proven in this paper state that the outgoing authority has a natural upper limit.

This upper limit does not depend on the inner structure of a cluster but rather on its boundaries.
So it may be an interesting further research direction to see to what extend this inner structure may influence getting closer or more far away from the theoretical limits.

As a further research direction it is also obvious that finding tighter limits is needed, or a proof that the found limits are the lowest ones possible. This would improve the evaluation of e.g.\  cluster quality.

In this paper we restricted ourselves to the case of preferential distribution of supernode authority among the nodes. A further research would be needed to cover the case of uniform distribution.

\subsection*{Acknowledgements}
The authors thank Polish Ministry of Science and Higher Education for supporting the research performed at our institute.

\bibliographystyle{fundam}
\bibliography{11345_bib}



\end{document}